\documentclass{procmult}
\usepackage{graphicx}
\usepackage{amsmath,amssymb,mathrsfs}
\usepackage{bbm}

\begin{document}

\title{Graviton propagator as a tool to test spinfoam models}
\author{Emanuele Alesci$^\dag$}
\institute{
$^\dag$Centre de Physique Th\'eorique de Luminy\footnote{Unit\'e mixte de recherche (UMR 6207) du CNRS et des Universit\'es
de Provence (Aix-Marseille I), de la M\'editerran\'ee (Aix-Marseille II) et du Sud (Toulon-Var); laboratoire affili\'e \`a la FRUMAM (FR 2291).}
, case 907, F-13288 Marseille, EU 
 \\
}
\maketitle
\abstract{I briefly review the advancements in the construction of the graviton propagator in the context of LQG and Spinfoam Models. In particular the problems of the Barrett-Crane vertex in giving the correct long-distance limit and the introduction of the new corrected models. This kind of calculation applied to an alternative vertex with given asymptotic can give the correct propagator and is then able to help selecting spinfoam models. In particular the study of the asymptotic properties of the new models shows the predicted behavior able to overcome the BC difficulties and to give the correct propagator}

\vspace{0.8cm}
Loop Quantum Gravity (LQG) \cite{lqg} is a promising candidate for a theory of Quantum Gravity based on a canonical quantization of General Relativity (GR). Its dynamics is defined through the action of the Hamiltonian constraint \cite{hamiltonian} but until now, the implementation of this constraint has encountered many difficulties that have led to path integral formulations of the same theory called Spinfoam Models (SM) \cite{lqg}. The most promising and extensive SM studied in the literature was the Barrett-Crane (BC) model \cite{Barrett:1997gw}. 
The main open problem is the contact with the low energy world namely on a way to recover from LQG and SM low energy predictions described by GR and its linearized quantization. 
Remarkably in the last few year a technique for computing $n$-point functions in this background-independent context has been introduced in \cite{scattering1, scattering3} and developed \cite{Livine:2006it} (see also \cite{simone} and references therein).
One of the main achievement of this technique has been the fact that this kind of calculation has ruled out the BC model \cite{I} and has lead to the introduction of a new class of spinfoam models. In fact the difficulties found in the construction of the propagator with the BC dynamics have focused the attention on the intertwiner dependence of these models (associated to the angles between simplices) revealing a strong imposition of the simplicity constraints in the quantization procedure that has the effect of freezing some degrees of freedom.
This problem has been corrected in a new model now called EPR \cite{EPR} that has the remarkable feature of having boundary states compatible with the boundary states of LQG.
This vertex can be defined for general values of the Immirzi parameter $\gamma$ \cite{EPRL} and extended to the Lorentzian case \cite{L}. A similar vertex now called FK \cite{FK} has also been derived  using the coherent states techniques introduced by Livine and Speziale \cite{LS}. For $\gamma<1$ the two techniques yield exactly the same 
theory. These models can now be tested with the propagator technique and another complementary based on the wave packet propagation \cite{numerico1,numerico2,fusion} to understand their semiclassical limit  \cite{CF1,CF2,CF3,BW} and to construct Feynman rules for quantum gravity.
\vspace{-0.9cm}
\section{LQG propagator: BC model and the New models}
Given a regular 4-simplex with two boundary tetrahedra $n$ and $m$ centered at the points $x$ and $y$ we can define the projection of the euclidean graviton propagator $G^{\mu\nu\rho\sigma}(x,y)$ as ${\mathbf G} _{n,m}^{\scriptscriptstyle ij,kl}(L)=
G^{\mu\nu\rho\sigma}(x,y)(n^{\scriptscriptstyle(i)}_n)_\mu (n^{\scriptscriptstyle(j)}_n)_\nu (n^{\scriptscriptstyle(k)}_m)_\rho (n^{\scriptscriptstyle(l)}_m)_\sigma$, where the latin indexes label the five tetrahedra bounding the 4-simplex and $n_m^{\scriptscriptstyle(k)}$ is the normal one-form to the triangle bounding the tetrahedra $m$ and $k$,  in the hyperplane defined by $m$, and $L$ is the euclidean distance between $x$ and $y$. ${\mathbf G} _{n,m}^{\scriptscriptstyle ij,kl}(L)$  can be computed \cite{scattering1} in a background  independent context as 
\begin{equation}
{\mathbf G} _{{\mathbf q}\, n,m}^{\scriptscriptstyle ij,kl} = \langle W | \big(E^{\scriptscriptstyle(i)}_n \cdot E^{\scriptscriptstyle(j)}_n-n_n^{\scriptscriptstyle(i)}\cdot n_n^{\scriptscriptstyle(j)}\big)
\big(E^{\scriptscriptstyle(k)}_m  \cdot E^{\scriptscriptstyle(l)}_m-n_m^{\scriptscriptstyle(k)}\cdot n^{\scriptscriptstyle(l)}_m\big) |\Psi_{\mathbf q} \rangle.
\label{partenza1}
\end{equation}
for an appropriate $\mathbf q$. We refer to \eqref{partenza1} as the LQG graviton propagator.  The computation of this object is based on three ingredients:  the boundary functional $ \langle W |$,   the triad operator $E_n^{\scriptscriptstyle(i)}$ at the point $n$, contracted with $n_n^{\scriptscriptstyle(i)}$ and the boundary state  $|\Psi_{\mathbf q} \rangle$, picked on a given classical boundary (intrinsic and 
extrinsic)  geometry $\mathbf q$. 
The diagonal components ${\mathbf G} _{{\mathbf q}\, n,m}^{\scriptscriptstyle ii,kk}$
have been computed \cite{scattering3} using the BC model to codify the dynamics, the double grasping operators of LQG acting in the same directions (the LQG area operators) and  a vacuum state given by a gaussian superposition of spinnetwork states. In \cite{scattering3} was shown that at large distance, these components of \eqref{partenza1} agree with the ones of the conventional graviton propagator. 

The construction of the non diagonal terms \cite{I,II} with the same tecnique, required a change in each of the three ingredients, because the graviton operators $E^{\scriptscriptstyle(i)}_n \cdot E^{\scriptscriptstyle(j)}_n$ (associated with the 3d dihedral angles between the triangles in the tetrahedra $n$) called into play the dependence of the spinnetworks from the \emph {intertwiners} and in turns, the dependence of the boundary state and the vertex from these variables.
In particular the BC dynamics used to compute the diagonal terms has a trivial intertwiner dependence that appeared insufficient to deal with the non diagonal terms. To see how this works it is enough to calculate \eqref{partenza1} to first order in the GFT \cite{lqg,GFT} expansion, and in the limit in which the boundary surface is large.
Eq. \eqref{partenza1} receive the leading contribution for $W$ with support only on spin networks with  graph dual to a 4-simplex. If 
${\mathbf j}$ and ${\mathbf i}$
are, respectively, the ten spins and the five intertwiners that color this graph, then in this
approximation (\ref{partenza1}) reads
\begin{equation}
	{\mathbf G} _{{\mathbf q}\, n,m}^{\scriptscriptstyle ij,kl} 
	 = \sum{}_{{\mathbf j}, {\mathbf i}} \,W({\mathbf j}, {\mathbf i}) \big(E^{\scriptscriptstyle(i)}_n \cdot E^{\scriptscriptstyle(j)}_n-n_n^{\scriptscriptstyle(i)}\cdot n_n^{\scriptscriptstyle(j)}\big) 
	 \big(E^{\scriptscriptstyle(k)}_m  \cdot E^{\scriptscriptstyle(l)}_m-n_m^{\scriptscriptstyle(k)}\cdot n^{\scriptscriptstyle(l)}_m\big)
	 \Psi({\mathbf j}, {\mathbf i}).
\label{partenza2}
\end{equation}
This expression has to be calculated using the double grasping operators acting on the nodes of the 4-symplex spinnetwork derived in \cite{I} and a suitable choice of boundary state that we want picked on the classical geometry $\mathbf q$ of a regular 4 simplex. The boundary state was defined \cite{scattering3,I,II} as 
gaussian wave packet in spin and intertwiners variables, centered on the values  $j_0$ and $i_0$ determined respectively by the areas and the angles of the background 4-simplex. As in ordinary quantum mechanics a crucial role was played by the phase of the packet. The spin phase coefficients were fixed by the background extrinsic geometry \cite{scattering3}. The intertwiner phase coefficients were fixed by the requirement that the state remain peaked after a change of pairing to the value $i_0$ determined by the classical value of the 3d dihedral angles of each regular tetrahedron inside the 4-simplex.  
The crucial point is that the non commutativity \cite{tetraedro} of the the different angles of a tetrahedron, represented by the intertwiner variables in different pairings, requires a state with a phase dependence in the intertwiner variables to be peaked on the background angles in any pairing.
The correct value  \cite{tetraedro2,I} for this in the equilateral case is $exp\{i\frac{\pi}{2}i_n \}$.
In \cite{scattering3} and \cite{I}, (a suitable adjustment of) the BC vertex was chosen
for $W$ and in this limit the propagator depends only on its asymptotic behavior, 
this has the structure \cite{asimp}
$
W_{BC}(\mathbf{j})\sim
		e^{\frac{i}{2}(\delta{\mathbf{j}} G \delta {\mathbf{j}})} e^{i\Phi \cdot \delta{\mathbf{j}}} +e^{-\frac{i}{2}(\delta {\mathbf{j}}G \delta {\mathbf{j}})} e^{-i\Phi\cdot \delta {\mathbf{j}}}
$
where $G$ is the $10\times10$ matrix given by the second derivatives of the 4d Regge action around
the symmetric state, and $\Phi$ is a 10d vector with all equal components, which were shown \cite{scattering3} to precisely match those determined by the background extrinsic curvature.
Computing \eqref{partenza2} with these ingredients the crucial point is that the phase in the link variable in the boundary state cancels with the phase of one of the two terms of $W_{BC}$, while 
the other term is suppressed \cite{scattering3} for large $j_0$ (this was the key mechanism that gave the correct diagonal components with the BC vertex) but \emph{the rapidly oscillating factor in the intertwiners variables is completely uncompensed by the dynamics and suppress the integral}\cite{I}. The intertwiner independence of the BC vertex prevents the propagator to have the correct long distance behavior.
In Ref. \cite {II} has been proposed an asymptotic form of a vertex $W$
 that includes a gaussian in all the 15 variables, and most \emph{crucially} a phase dependence also on the intertwiner variables. 
The proposed form for the asymptotic of $W$ was
	$W(\mathbf{j},\mathbf{i}) = 
	e^{\frac{i}{2}(\delta{\mathbf{I}} G \delta {\mathbf{I}})} e^{i\phi \cdot \delta{\mathbf{I}}} +e^{-\frac{i}{2}(\delta {\mathbf{I}} G \delta {\mathbf{I}})} e^{-i\phi\cdot \delta {\mathbf{I}}}
$  , where $G$ is a $15\times15$ real matrix that scales as $1/j_0$.  The quantity $\phi=(\phi_{nm},\phi_n)$ is now a 15d vector: 
its 10 spin components
$\phi_{nm}$ just reproduce the spin phase dependence of the BC vertex; while its five intertwiner components are equal and fixed to the value 
$\phi_n=\frac{\pi}{2}$.
This phase dependence is the crucial detail, that makes the calculation work because it allows the cancellation of the phases between the propagation kernel and the boundary state through which the dynamical kernel reproduces the 
semiclassical dynamics in quantum mechanics. If
this does not happens, the rapidly oscillating phases suppress the amplitude. 
The results of Ref. \cite{I} motivated the search for alternative models.
Ref. \cite{II} shows that it is possible to recover the full propagator of the linearized theory from the LQG propagator and gives indications on the asympthotic behavior of an alternative vertex: in particular it requires for the new models an oscillation in the intertwiners. A natural question was then if the new models present or not the desired oscillation.

Let us focus on the EPRL model \cite{EPRL}. For given Immirzi parameter $\gamma$, the vertex amplitude is defined as a function of five $SO(3)$ intertwiners $i_n$ and ten spins $j_{nm}$ 
\begin{equation}
	W(j_{nm},i_n)=\sum_{i_n^L\,i_n^R}\, \{15j\}_N\big(\frac{|1-\gamma|j_{nm}}{2},i_n^L\big)\;\, \{15j\}_N\big(\frac{(1+\gamma)j_{nm}}{2},i_n^R\big)\;
	\prod_n  f^{i_n}_{i_n^L i_n^R}(j_{nm})\;.
	\label{eq:EPRL}
\end{equation}
The functions $\{15j\}_N$ are normalized $15j$-symbols and the $f^{i_n}_{i_n^L i_n^R}$ are fusion coefficients from $SO(3)$ to $SU(2)_L\times SU(2)_R$ introduced in \cite{EPR}. Indeed the model differs from the BC one only for the structure of these coefficients.
One of the main open question on these new models was 
their large large spin's asymptotics to repeat the propagator calculations, with particular focus on the new intertwiner dependence, and their semiclassical limit to establish a link with Regge calculus.
The analysis of these models started numerically using a new technique to test the spinfoam dynamics introduced in \cite{numerico1}: The propagation of semiclassical wavepackets. 
This technique is based on the following simple consideration:
In ordinary quantum mechanics the propagation kernel $ W_t(x,y)=\langle x|e^{-\frac \imath\hbar Ht} |y\rangle
$ of a one-dimensional nonrelativistic quantum system defined by a hamiltonian operator $H$  can be used to check the semiclassical behaviour of the theory using the propagation of wave packets $\psi_{x,p}(x)$ with $W_t(x,y)$. If the propagation is coherent, given a semiclassical wave packet $\psi_{x_i,p_i}(y)$  
centered on the initial values $x_i, p_i$ its evolution under the 
kernel $
    \phi(x):=\int dy\  W_t(x,y)\ \psi_{x_i,p_i}(y)   
$
will be a new semiclassical wave packet centered around the final data  $x_f, p_f$.
Now the equations of motion of any dynamical system can
be expressed as constraints on the set formed by the initial, final variables namely on the boundary variables. This means that a solution of Einstein equations can be recovered by an appropriate set of constraints on the boundary variables.
The derivation of the vertex amplitude presented in
\cite{EPR} indicated that the process described by one vertex
can be seen as the dynamics of a single cell in a Regge
triangulation of general relativity.   
It follows that a boundary wave packet centered on the boundary variables of a flat 4-symplex must be correctly propagated by the vertex amplitude evolving in a wave packet centered on the correct boundary variables, if the vertex amplitude is to give the Einstein equations in the
classical limit.
The tested process was the evolution of four  identical coherent tetrahedra \cite{tetraedro2} with the propagation kernel of the EPR model.
A single regular Regge cell has all the boundary variables equals: this means that if the quantum kernel describes the same process, the final state should have been a coherent tetrahedron with the same geometrical properties of the evolved states. In \cite{numerico1} the expected mean and phase were found and in \cite{numerico2} were found the correct analytical properties of the evolved state. 
To calculate this propagation were used the asymptotic properties of the fusion coefficient \cite{fusion} that allowed drastic simplifications in the numerical calculations and a new numerical algorithm (using techniques similar to \cite{Khavkine:2008kk}). The asymptotic properties of fusion coefficients showed that these coefficients not only give nontrivial dynamics to intertwiners at the quantum level, but also map semiclassical SO(3) tetrahedra into semiclassical SO(4) tetrahedra. 
To complete the analysis of the asymptotic properties of the model with this approach it is needed the asymptotic formula for the $15j$-symbol that is still missing. 
Though was used the drastic approximation of fixing all spins, these were the first indication that the EPR model has the good semiclassical limit and the propagation scheme gave an hint on the required linear phase dependence in the asymptotic regime (in fact the phase of the outgoing state was exactly $\frac{\pi}{2}$ and this phase comes directly from the vertex).
In the meanwhile dealing with the tentative of relating the spinfoam models to a physical scalar product a new integral formulation has been given \cite{Alesci:2008yf}.

Key developments appeared in \cite{CF2} were the authors  studied the semiclassical properties of the FK model using the construction of a path integral with a discrete and local action introduced in \cite{CF1}. They showed that, in the semi-classical limit, the amplitude converges rapidly towards the exponential of $i\; S_{Regge}$, if the face's spins can be understood as coming from a discrete geometry, otherwise the spin foam amplitude is exponentially suppressed. Remarkably this result holds for an arbitrary triangulation and gave important informations on the good semiclassical limit of the model and its relation with Regge calculus, but it was not yet viable for the explicit calculation of the graviton propagator, where the explicit dependence on the intertwiners had to be understood.

A new crucial result is the semiclassical limit recently performed by Barrett, Fairbairn and collaborators \cite{BW} and by Conrady, Freidel \cite{CF3} that gives the exponential of the Regge action for the vertex projected on the Livine-Speziale coherent state basis.
This important result is now directly applicable in the propagator calculations. In \cite{mio3} it is noted that this result implies for the equilateral configuration
$ W_{EPR}(j_0,i_0+\delta \mathbf{i}) \sim N e^{-\imath S(\mathbf{j_0})}\  e^{i \frac{\pi}2 i}.
$
The important point here is the appearance of the correct $\frac{\pi}{2}$ factor in the phase, which was missing in the BC vertex.  Thus, the new vertex has the asymptotic behavior that was \emph{guessed} in \cite{II}, in order to yield the graviton propagator. The obstacle that prevented the BC vertex to yield the proper propagator is resolved by the new vertex. 
The complete calculation with the new vertex is now possible \cite{noi} and in course.

\end{document}